\documentclass[12pt]{iopart}

\usepackage{iopams}

\expandafter\let\csname equation*\endcsname\relax

\expandafter\let\csname endequation*\endcsname\relax

\usepackage{amsmath}
\usepackage{graphicx}
\usepackage{bm}
\usepackage{amssymb}
\usepackage{float}
\usepackage{arydshln}
\usepackage[dvips]{color}

\newcommand{\NRB} {NbRh$_{2}$B$_{2}$}
\newcommand{\TRB} {TaRh$_{2}$B$_{2}$}
\newcommand{\TC} {$T_{\mathrm{c}}$}

\begin{document}

\title{Superconductivity and the upper critical field in the chiral noncentrosymmetric superconductor NbRh$_{2}$B$_{2}$}

\author{D. A. Mayoh$^1$, M. J. Pearce$^1$, K. G\"{o}tze$^1$,  A. D. Hillier$^2$, G. Balakrishnan$^1$, M. R. Lees$^1$}

\address{$^1$ Physics Department, University of Warwick, Coventry, CV4 7AL, United Kingdom}
\address{$^2$ ISIS facility, STFC Rutherford Appleton Laboratory, Harwell Science and Innovation Campus, Oxfordshire OX11 0QX, United Kingdom}
\ead{D.Mayoh@warwick.ac.uk}
\ead{M.R.Lees@warwick.ac.uk}
\vspace{10pt}
\begin{indented}
\item[]\today
\end{indented}

\begin{abstract}
\NRB\ crystallises in a chiral noncentrosymmetric structure and exhibits bulk type-II superconductivity below $7.46(5)$~K. Here we show that the temperature dependence of the upper critical field deviates from the behaviour expected for both Werthamer-Helfand-Hohenberg and the Ginzburg-Landau models and that $\mu_{0}H_{\mathrm{c2}}\left(0\right)\approx 18$~T exceeds the Pauli paramagnetic limit, $\mu_{0}H_{\mathrm{P}}=13.9$~T. We explore the reasons for this enhancement.  Transverse-field muon spectroscopy measurements suggest that the superconducting gap is either $s$-wave or $\left(s+s\right)$-wave, and the pressure dependence of \TC\ reveals the superconducting gap is primarily $s$-wave in character. The magnetic penetration depth $\lambda(0) = 595(5)$~nm. Heat capacity measurements reveal the presence of a multigap $\left(s+s\right)$-wave superconducting order parameter and moderate electron-phonon coupling.
\end{abstract}

%
%
%
%

\section{Introduction}
\label{Introduction}

The observation of multigap superconductivity in MgB$_2$ renewed interest in the boride based superconductors~\cite{Bouquet01, Liu, Iavarone, Zehetmayer} as potential hosts of high-temperature superconductivity. At the same time, studies of noncentrosymmetric (NCS) superconductors have demonstrated how the structure of a compound can have a profound effect on the superconducting properties of a material~\cite{Smidman17, Bauer12}. The recent discovery of the chiral noncentrosymmetric superconductors \NRB\ and \TRB, provides a new platform for exploring the relationship between superconductivity and structure~\cite{Carnicom, Mayoh}.

A noncentrosymmetric material has a crystal structure that lacks a centre of inversion. The asymmetric positioning of atoms within the unit cell gives rise to local electric-field gradients and asymmetric spin-orbit coupling (ASOC). This ASOC can lift the spin degeneracy of the electronic bands, splitting the Fermi surface into spin-polarised subsurfaces. In a NCS superconductor below its transition temperature, \TC, the Cooper pairs may then form in a parity mixed state~\cite{Bauer12} although typically, one channel is dominant over the other~\cite{Smidman17, Bauer12}. An admixture of singlet and triplet pairs can, in turn, give rise to exotic superconducting gap structures, magnetoelectric effects such as upper critical fields that exceed the Pauli limit, and time-reversal symmetry breaking~\cite{LaNiC2, Re6Zra, La7Ir3}.

Initially, research on NCS superconductors focused on heavy-fermions such as CePt$_{3}$Si~\cite{Bauer04}, CeRhSi$_{3}$~\cite{Kimura07}, CeIrSi$_{3}$~\cite{Sugitani06}, and CeCoGe$_{3}$~\cite{Settai07} because of their unusual superconducting gap structures and upper critical fields that lie well above the Pauli limit. Later, however, it became clear that it is necessary to study noncentrosymmetric systems without heavy-fermion features, as the strongly correlated nature of $f$ electrons may obscure any behaviour associated with singlet-triplet mixing. Many weakly correlated noncentrosymmetric compounds have now been characterised, and while the majority exhibit a fully gapped $s$-wave superconductivity, a number of fascinating superconductors have been discovered. 


Li$_{2}$(Pd$_{1-x}$Pt$_x$)$_{3}$B $\left(0 \leq x\leq 1\right)$, is an example of a weakly correlated superconductor that crystallises with an antiperovskite, NCS, chiral structure (space group $P4_332$). It has been reported that replacing Pd with Pt leads to an evolution from an isotropic $s$-wave gap in Li$_{2}$Pd$_{3}$B to a line-nodal gap in Li$_{2}$Pt$_{3}$B. It is thought that the stronger ASOC arising from the Pt enhances the triplet component of the superconductivity in Li$_{2}$Pt$_{3}$B~\cite{Yuan2006, Nishiyama2007}. The strongly coupled superconductor Mo$_{3}$Al$_{2}$C, (space group $P4_132$), also has NCS chiral structure. A nodal gap structure has been reported in Mo$_{3}$Al$_{2}$C from heat capacity and nuclear magnetic resonance (NMR) measurements~\cite{Mo3Al2C, Bauer10}. Under pressure, the \TC\ of Mo$_{3}$Al$_{2}$C is seen to increase, providing further evidence of singlet-triplet mixing~\cite{Bauer10}. 


While an admixture of singlet and triplet states may be expected to lead to a nodal gap structure, other unconventional gap structures are also found in NCS superconductors, and these too can lead to interesting new physics. For example, Y$_2$C$_3$ is a noncentrosymmetric superconductor with a high upper critical field; $H_{\mathrm{c2}}\left(0\right)=27$~T which exceeds the weak-coupling Pauli limit. Both NMR~\cite{Harada2007} and muon spectroscopy~\cite{Kuroiwa08} suggest that Y$_2$C$_3$ has a two-band, $\left(s+s\right)$-wave symmetry with inter-band coupling. On the other hand, tunnel diode measurements of the penetration depth suggest that while a two-gap model can describe the behaviour close the \TC, the lower temperature behaviour is indicative of line nodes in the superconducting energy gap~\cite{Chen11}.


Multigap superconductivity has also been found to occur in the iron pnictide superconductors~\cite{Khasanov15, Stewart11, Si16} as well as the borides, including MgB$_2$~\cite{Bouquet01, Liu, Iavarone, Zehetmayer}. A combination of disorder and strong spin-orbit coupling can lead centrosymmetric materials to develop unusual superconducting properties. $R_2$Pd$_x$S$_5$ $\left(x\leq 1\right)$ where $R=$~Nb or Ta form a remarkable family of centrosymmetric compounds~\cite{Zhang13, Lu14, Biswas15}. In Ta$_{2}$Pd$_{x}$S$_{5}$  ($T_{\mathrm{c}}\sim 6$~K) disorder due to a Pd deficiency is introduced into the compound, giving rise to Anderson localisation. This along with strong SO coupling, leads to a highly anisotropic upper critical field, with an $H_{\mathrm{c2}}\left(0\right)$ along the $b$ axis of 37~T exceeding the Pauli limit~\cite{Lu14}.

\NRB\ is one member of a new family of NCS superconductors with a chiral structure. The first report on this material~\cite{Carnicom} showed that \NRB\ crystallises in a trigonal structure with the space group $P3_1$. The \TC\ of \NRB\ was reported to be 7.5~K and an unusually large upper critical field that exceeds the Pauli limit was also observed. Here, we have used a combination of muon spectroscopy, heat capacity, magnetisation, and resistivity measurements to probe the nature of the superconducting state of \NRB.  We present heat capacity data that can best be described by a two-gap $\left(s+s\right)$-wave model. Transverse-field $\mu$SR data show that the superconducting gap can be described by an isotropic $s$-wave model, but is also compatible with an isotropic multigap $\left(s+s\right)$-wave model, while a decrease in \TC\ with pressure suggests that the superconducting gap is dominated by an $s$-wave component. We also present an extended $H-T$ phase diagram of \NRB\ with upper critical field measurements up to 17~T. Both the Werthamer-Helfand-Hohenberg (WHH) model and the Ginzburg-Landau (GL) model produce poor fits to the upper critical field of \NRB\ at lower temperatures and higher fields. We postulate that a two-gap model may be able to describe the $H_{\mathrm{c2}}\left(T\right)$ behaviour, but further investigations of the inter- and intra-band scattering are required to confirm this suggestion.

\section{Experimental Details}
\label{Experimental Details}

Polycrystalline samples of \NRB\ were prepared from the constituent elements using the synthesis method described earlier~\cite{Carnicom}. Powder X-ray diffraction (PXRD) performed using a Panalytical X-Pert Pro diffractometer showed the material forms with a trigonal $P3_1$ structure in agreement with earlier reports~\cite{Carnicom}. The samples were confirmed to be made up of a single phase to within the detection limit of the technique, but with a single unidentified Bragg peak at a $d$ spacing of 2.1~\AA. 

DC magnetisation, $\left(M\right)$, measurements as a function of temperature, $\left(T\right)$, at a fixed field, or as a function of applied magnetic field, $\left(H\right)$, at a fixed temperature, were made in a Quantum Design Magnetic Property Measurement System (MPMS) magnetometer at temperatures between 1.8 and 300~K and in magnetic fields up to 5~T. Magnetisation versus applied field data were also collected in an Oxford Instruments vibrating sample magnetometer (VSM) in magnetic fields up to 7~T. The ac susceptibility, $\left(\chi_{\mathrm{ac}}\right)$, measurements were performed in a Quantum Design MPMS with an ac applied field of 0.3~mT and a frequency of 30~Hz in dc magnetic fields up to 5~T. Magnetisation measurements under external hydrostatic pressure, $\left(P\right)$, up to 10~kbar where performed using an Almax easyLab BeCu Mcell 10 pressure cell in a MPMS magnetometer. The shift in the superconducting transition temperature of tin was used as a manometer to determine the value of the applied pressure. A Quantum Design Physical Property Measurement System (PPMS) with a $^3$He insert was used to measure the resistivity, $\left(\rho\right)$, as a function of temperature at fixed field, or as a function of applied magnetic field at a fixed temperature, from 0.5 to 300~K in fields up to 9~T. Resistivity measurements in magnetic fields of up to 17~T were made using an Oxford Instruments cryomagnet. Silver wires were fixed to the sample in a four-point geometry using DuPont 4929N silver paste. Measurements were made using ac or square-wave currents. Heat capacity, $\left(C\right)$, measurements were also carried out in a QD PPMS with a $^3$He and a dilution fridge insert from 0.1 to 300~K in magnetic fields up to 9~T. The sample was attached to a sapphire stage using Apiezon N-grease. Measurements of the heat capacity of the empty stage and the stage and sample together were made using a two-tau relaxation method.   

Zero-field (ZF) and transverse-field (TF) muon spin relaxation/rotation, ($\mu$SR), measurements were made at the ISIS pulsed muon facility using the MuSR spectrometer.  Muons produced in pulses were implanted into the sample where each muon decayed with a half-life of 2.2~$\mu$s into a positron and two neutrinos. The positrons were captured in 64 detectors positioned around the sample allowing the asymmetry in the positron emission to be measured as a function of time. A description of the detector geometries is given in Ref.~\cite{SLee}. The polycrystalline sample of NbRh$_{2}$B$_{2}$ was fixed to a 99.995\% silver plate using GE varnish diluted with ethanol and then loaded into a $^3$He cryostat allowing data to be collected between 300~mK and 8.5~K. For the measurements in zero field, stray fields at the sample position were cancelled to within 1~$\mu$T by an active compensation system. For the transverse-field measurements, the sample was cooled in applied magnetic fields of up to 40~mT and the data recorded on warming.

\section{Magnetic Properties and Lower Critical Fields}
\label{Magnetic}

\begin{figure}[tb]
\centering
\includegraphics[width=0.45\columnwidth]{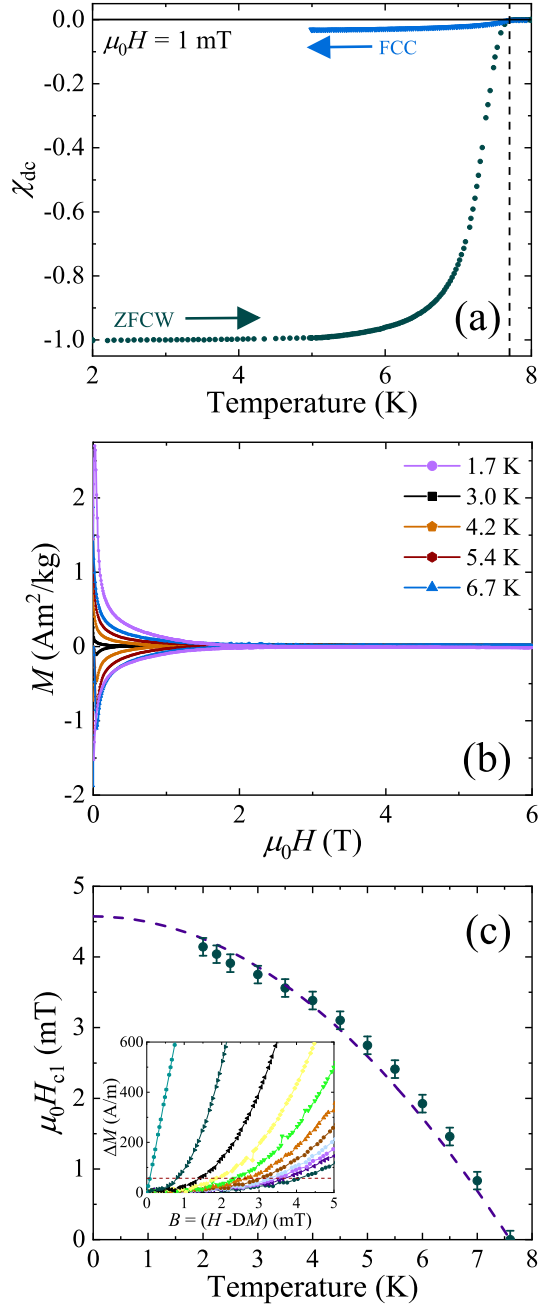}
\caption{(Colour online) (a) Temperature dependence of the dc magnetic susceptibility, $\chi_{\mathrm{dc}}\left(T\right)$, collected in zero-field-cooled warming (ZFCW) and field-cooled cooling (FCC) mode in an applied field of $\mu_{0}H = 1$~mT. (b) Magnetisation versus magnetic field at several temperatures for \NRB\ exhibits a behaviour typical for a type II superconductor. The data were collected in a VSM with the demagnetisation factor of the sample minimised. (c) Lower critical field, $H_{\mathrm{c1}}$, versus temperature for \NRB. The $H_{\mathrm{c1}}$ values were taken to be the fields at which the magnetisation versus field data first deviate from linearity. The dashed line shows the fit using Eq.~\ref{hc1} giving $\mu_{0}H_{\mathrm{c1}}(0) = 4.6(1)$~mT. The inset shows the demagnetisation corrected residuals for linear fit to $M$ versus $H$ at several temperatures.}
\label{Magnetisation}
\end{figure}

\begin{figure}[tb]
\centering
\includegraphics[width=0.5\columnwidth]{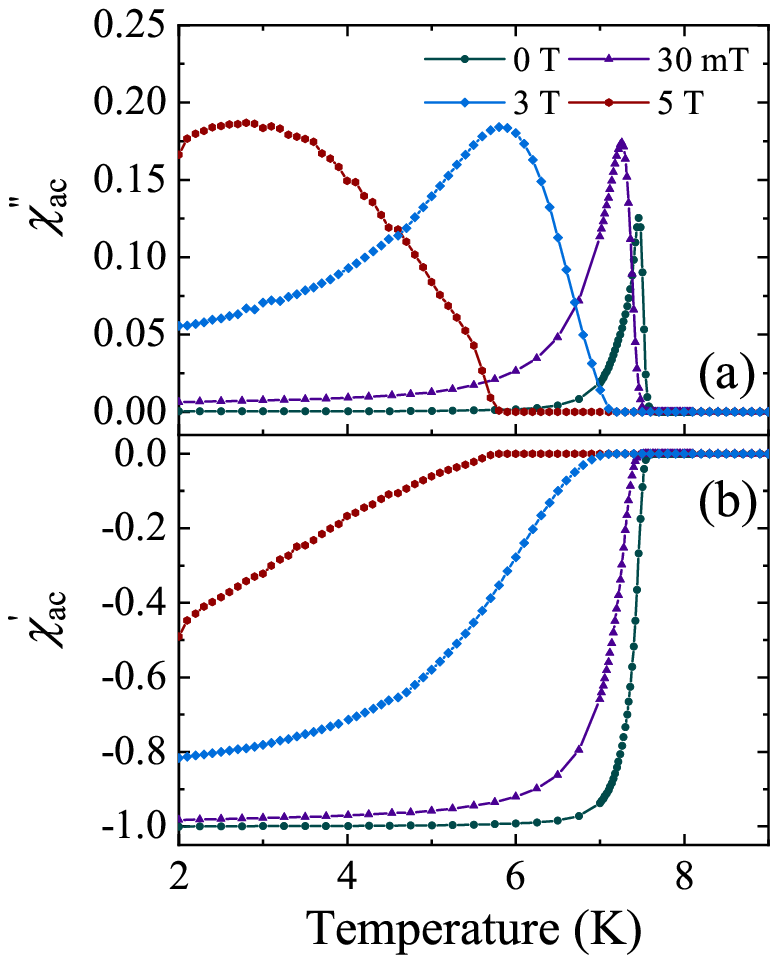}
\caption{(Colour online) Temperature dependence of (a) the imaginary part of ac susceptibility, $\chi''\left(T\right)$, and (b) the real part of ac susceptibility, $\chi'\left(T\right)$ for \NRB\ in dc applied fields of up to 5~T. A sharp superconducting transition can be seen at $7.50\left(5\right)$~K in zero dc field. In dc fields above $H_{\mathrm{c1}}$ the transition broadens slightly and shifts to lower $T$. $\chi''\left(T\right)$ indicates that there is considerable vortex motion at higher fields.}
\label{ChiT}
\end{figure}

A polycrystalline sample of \NRB\ was first characterised by studying the temperature dependence of the dc magnetic susceptibility $\chi_{\mathrm{dc}}\left(T\right)$. Zero-field-cooled warming (ZFCW) and field-cooled cooling (FCC) data collected in an applied field of 1~mT are shown in Fig.~\ref{Magnetisation}(a). The onset of the superconducting state in \NRB\ is observed at $T^{\mathrm{onset}}_{\mathrm{c}} = 7.58(5)$~K which agrees with a previous report~\cite{Carnicom}. A full Meissner fraction is observed at $5$~K in the ZFCW data indicating complete flux expulsion and bulk superconductivity in the sample. The FCC data show a small amount of flux expulsion indicating moderate to weak pinning of the magnetic field within \NRB\ in the superconducting state. 

The field dependence of the dc magnetisation for \NRB\ was measured at several temperatures between 1.7 and 6.7~K as shown in Fig.~\ref{Magnetisation}(b). The magnetic hysteresis loops show behaviour typical of a type-II superconductor. The $M\left(H\right)$ loops are symmetric about $M=0$ suggesting bulk pinning plays a dominant role in determining the overall form of the $M\left(H\right)$ data. At higher fields and lower temperatures there are discontinuities in the data due to flux jumps. These flux jumps disappears at low fields and are not present in the data-set collected at 6.7 K. The hysteresis in $M\left(H\right)$ initially decreases rapidly with increasing field, before a much weaker trend is established at higher fields. The hysteresis in $M\left(H\right)$ persists up to fields of the order of a tesla, ($\sim 4$~T at 1.7~K), and then vanishes (below the resolution of the VSM). For higher field the magnetisation becomes reversible. This occurs at temperatures and magnetic fields well below the $H_{\mathrm{c2}}\left(T\right)$ values determined from the resistivity and the heat capacity measurements (see below). The width of the magnetic hysteresis, $\Delta M$, can be related to the critical current density, $J_{\mathrm{c}}$, via the Bean critical state model~\cite{Bean64}. Therefore, in the reversible regime, where the vortex lattice is free to move and can be thought of as liquid, the material may no longer carry a finite bulk supercurrent. These results support the view that this sample of \NRB\ exhibits moderate to weak bulk pinning of the vortex lattice. 

Figure~\ref{Magnetisation}(c) shows the lower critical field, $H_{\mathrm{c1}}$, as a function of temperature. The lower critical fields were approximated by determining the point at which the $M\left(H\right)$ data first deviate from linearity ($\Delta M=90$~A/m) in the low-field magnetisation loops at each temperature, as shown in the inset of Fig.~\ref{Magnetisation}(c). The data in Fig.~\ref{Magnetisation}(c) were fit with the Ginzburg-Landau relation
\begin{equation}
H_{\mathrm{c1}}\left(T\right) = H_{\mathrm{c1}}\left(0\right)\left[1 - \left(\frac{T}{T_{\mathrm{c}}}\right)^{2}\right],
\label{hc1}
\end{equation}
giving a value of $\mu_{0}H_{\mathrm{c1}}\left(0\right) = 4.6(1)$~mT. This value is significantly lower than the 13.5~mT reported in Ref.~\cite{Carnicom}.

The temperature dependence of the ac magnetic susceptibility, $\chi_{\mathrm{ac}}(T)$, in several dc applied fields is shown in Fig.~\ref{ChiT}. The data in zero dc field confirm the superconducting onset temperature of $T^{\mathrm{onset}}_{\mathrm{c}} = 7.58(5)$~K. The \TC\ of \NRB\ is only slowly suppressed in field, falling to 5.7(1)~K in an dc field of $\mu_{0}H=5$~T. In zero dc field the sample exhibits a full Meissner fraction. For dc bias fields of the order of $H_{\mathrm{c1}}\left(0\right)$, the out-of-phase component of the ac susceptibility, $\chi''(T)$, has a sharp peak close to \TC\ which then falls to zero at lower-temperatures, indicating that the flux lines are pinned. In higher dc fields of $\mu_{0}H \geq 3$~T, a broad maximum is seen in $\chi''(T)$ typical of flux motion within the sample, and a partial flux penetration leads to a reduced Meissner fraction. Nevertheless, the in-phase component of the ac susceptibility, $\chi'(T)$, still exceeds 40\% in an applied field of $\mu_{0}H=5$~T.

\section{$\mu$SR measurements}
\label{muSR}

\begin{figure}[tb]
\centering
\includegraphics[width=0.5\columnwidth]{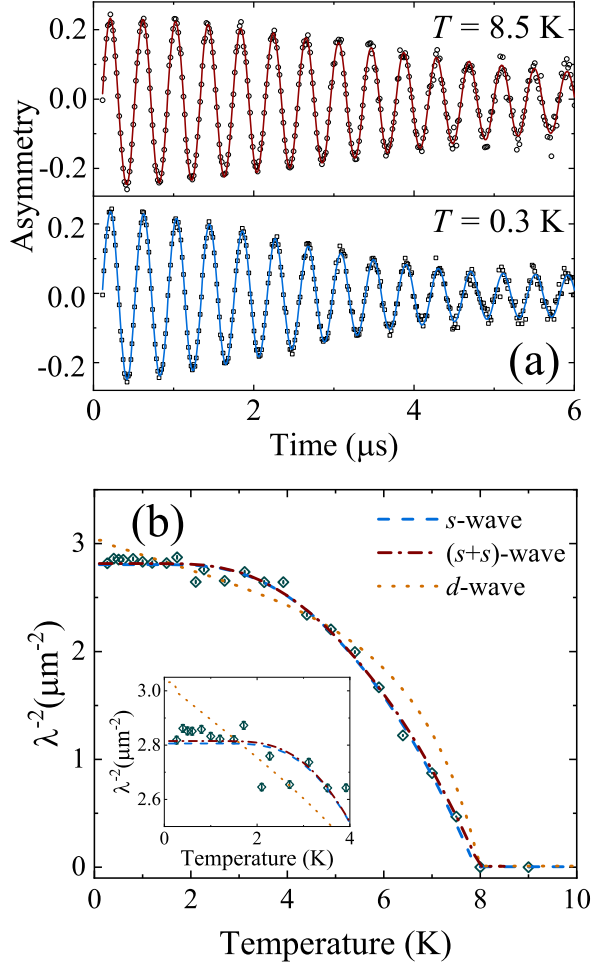}
\caption{(Colour online) (a) Time-dependent transverse-field $\mu$SR spectra for \NRB\ collected in the normal and superconducting state. The spectra were fit using Eq.~\ref{TF1} (solid lines) which models the Gaussian relaxation of the oscillatory signal due to the effects of the flux-line lattice. In the normal state the Gaussian relaxation observed below \TC\ is reduced but still exists due to a randomly oriented array of nuclear moments.  (b) Inverse square of the penetration depth, $\lambda^{-2}$, as a function of temperature for \NRB. The fits to the data using Eqs.~\ref{PenetrationDepth}~and~\ref{weighting} for the one- and two-gap models in the clean limit are shown by the lines. The inset shows the low-temperature data and fits on an expanded scale.}
\label{ZF_TFdata}
\end{figure}

In order to measure the structure of the superconducting gap in \NRB\ transverse-field $\mu$SR experiments were performed in applied fields of 18 and 30~mT, well above the $\mu_0H_{\mathrm{c1}}(0)\sim 4.6$~mT to ensure the sample is in the superconducting mixed state. The detectors in the spectrometer were grouped in eight blocks, each with a phase offset $\phi$, and the data were collected in field-cooled-warming (FCW) mode to ensure the most uniform flux-line lattice possible. Examples of the TF-$\mu$SR precession signals above and below \TC\ are shown in Fig.~\ref{ZF_TFdata}(a). The inhomogeneous field distribution can be modelled using a Gaussian relaxation function with a sinusoidal oscillating component~\cite{Khasanov15, SLee}. $G\left(t\right)$ also includes an oscillatory term that arises from muons which are directly implanted into the Ag sample holder and do not depolarise
\begin{equation}
G\left(t\right)=A_{1}\exp\left(-\frac{\sigma^2t^{2}}{2}\right)\cos\left(2\pi v_{1}t+\phi\right) +A_{2}\cos\left(2\pi v_{2}t+\phi\right),
\label{TF1}
\end{equation}
\noindent where $v_{1}$ and $v_{2}$ are the muon precession and background signal frequencies respectively and $\sigma$ is the Gaussian muon-spin relaxation rate. The total relaxation rate, $\sigma$, is given by a contribution from superconductivity in the sample, $\sigma_{sc}$, added in quadrature with a component due to the nuclear magnetic dipolar moments, $\sigma_{\mathrm{N}}$. 
\begin{equation}
\sigma_{\mathrm{sc}}=\sqrt{\left(\sigma^2-\sigma_{\mathrm{N}}^2\right)}.
\label{quad}
\end{equation}

\noindent Using Eq.~\ref{quad}, $\sigma_{sc}$ is calculated where $\sigma_{\mathrm{N}}$ was measured above \TC\ and assumed be constant over the temperature range studied. For a type-II superconductor where the applied field is much less than the upper critical field, $\sigma_{\mathrm{sc}}$ is directly related to the magnetic penetration depth, $\left(\lambda\right)$, by the expression 
\begin{equation}
\frac{\sigma^{2}_{\mathrm{sc}}\left(T\right)}{\gamma^{2}_{\mu}} = 0.00371 \frac{\Phi^{2}_{0}}{\lambda^{4}\left(T\right)},
\end{equation}\noindent
where {${\Phi_{0} = 2.068 \times 10^{-15}~\mathrm{Wb}}$} is the magnetic-flux quantum and {${\gamma_{\mu}/2\pi = 135.5~\mathrm{MHz/T}}$} is the muon gyromagnetic ratio~\cite{Brandt03, Sonier00}. The superfluid density is directly related to $\lambda$ so can be used to determine the superconducting gap structure in \NRB. In the clean and dirty limit, the temperature dependence of the magnetic penetration depth, $\lambda\left(T\right)$, can be calculated using

\begin{equation}
\label{PenetrationDepth}
\left[\frac{\lambda^{-2}\left(T,\Delta_{0}\right)}{\lambda^{-2}\left(0,\Delta_{0}\right)}\right]_{\mathrm{clean}}=1+\frac{1}{\pi}\int^{2\pi}_{0}\int^{\infty}_{\Delta_{\left(T,\phi\right)}}\left(\frac{\partial f}{\partial E}\right)\frac{EdE~d\phi}{\sqrt{E^2-\Delta\left(T,\phi\right)^2}},
\end{equation}
and
\begin{equation}
\label{PenetrationDepth_dirty}
\left[\frac{\lambda^{-2}\left(T,\Delta_{0}\right)}{\lambda^{-2}\left(0,\Delta_{0}\right)}\right]_{\mathrm{dirty}}=\frac{\Delta\left(T\right)}{\Delta\left(0\right)}\tanh\left[\frac{\Delta\left(T\right)}{2 k_{\mathrm{B}}T}\right],
\end{equation}
\noindent respectively, where $f=\left[1+\exp\left(E/k_BT\right)\right]^{-1}$ is the Fermi-Dirac function, and the temperature and angular dependence of the gap is given by $\Delta\left(T,\phi\right)=\Delta_{0}\delta\left(T/T_c\right)g\left(\phi\right)$. For an $s$-wave gap, the angular dependence of the superconducting gap function $g\left(\phi\right) = 1$, while for a $d$-wave gap $g\left(\phi\right) = \lvert \cos\left(2\phi\right)\rvert$, where $\phi$ is the azimuthal angle along the Fermi surface~\cite{Tinkham, Prozorov06}. $\Delta_0$ is the gap magnitude at zero kelvin, $\Delta\left(0\right)$, and the temperature dependence of the gap can be approximated by $\delta\left(T/T_c\right)=\tanh\left\{1.82\left[1.018\left(T_c/T-1\right)\right]^{0.51}\right\}$~\cite{Carrington}. Multigap analysis~\cite{Kuroiwa08, Khasanov15, Biswas15, BiswasLu2fe3Si5} was performed using a weighted sum of two superconducting gaps:

\begin{equation}
\label{weighting}
\left[\frac{\lambda^{-2}\left(T\right)}{\lambda^{-2}\left(0\right)}\right] =w \left[\frac{\lambda^{-2}\left(T,\Delta_{0,1}\right)}{\lambda^{-2}\left(0,\Delta_{0,1}\right)}\right] + \left(1-w\right)\left[\frac{\lambda^{-2}\left(T,\Delta_{0,2}\right)}{\lambda^{-2}\left(0,\Delta_{0,2}\right)}\right].
\end{equation}

\noindent The procedure followed here, although based on standard Bardeen-Cooper-Schrieffer (BCS) theory, has been widely used to study the superconducting properties of systems which do not fulfil all the strict criteria of a BCS superconductor and has played an important role in providing evidence for two-gap superconductivity in various materials including boride and iron-based superconductors.

The temperature dependence of the inverse penetration depth squared $\lambda^{-2}\left(T\right)$ is shown in Fig.~\ref{ZF_TFdata}(b). Both an isotropic $s$-wave model and an isotropic two-gap $\left(s+s\right)$-wave model in the clean and the dirty limits produce good fits to the data (see Table~\ref{MuonFits}). The gap values are slightly higher in the clean limit. For clarity only the clean limit fits are shown in Fig.~\ref{ZF_TFdata}(b).  Attempts were also made to fit $\lambda^{-2}\left(T\right)$ using a $d$-wave and an $\left(s+d\right)$-wave model in the clean limit. The $d$-wave produced a poor fit to the data and the fit using the $\left(s+d\right)$-wave model did not converge. There is little difference between the $\chi^2_{\mathrm{norm}}$ produced by the $\left(s+s\right)$- and $s$-wave models in both the clean and dirty limits, although on this basis alone an $\left(s+s\right)$-wave model would be slightly preferred. However, the uncertainty associated with both the magnitude of the smaller gap and the weighting, $w$, means a single-gap $s$-wave model remains a possibility. The value of $\Delta_{0}/k_{\mathrm{B}}T_{\mathrm{c}}=2.19(6)$ for the $s$-wave model, and $\Delta_{0, 1}/k_{\mathrm{B}}T_{\mathrm{c}}=2.3(2)$ for the two-gap $\left(s+s\right)$-model (both the clean limit), exceed the value expected for a BCS superconductor, $\Delta_\mathrm{{BCS}}/k_{\mathrm{B}}T_{\mathrm{c}} = 1.76$, which suggests that there is an enhancement in the electron-phonon coupling.

\begin{table}[tb]
\centering
\begin{tabular}{c c c c c}
\hline\hline
Clean limit &&&& \\
\hline
Gap & $\Delta_{0, 1}/k_{\mathrm{B}}T_{\mathrm{c}}$ & $\Delta_{0, 2}/k_{\mathrm{B}}T_{\mathrm{c}}$ & $w$ & $\chi^2_{\mathrm{norm}}$\\\hdashline
$s$-wave & 2.19(6) &  - & 1  & 1.95\\ 
$\left(s+s\right)$-wave & 2.3(2)  & 0.8(8) & 0.93(7) & 1.87 \\  
\hline 
Dirty limit &&&& \\
\hline 
Gap & $\Delta_{0, 1}/k_{\mathrm{B}}T_{\mathrm{c}}$ & $\Delta_{0, 2}/k_{\mathrm{B}}T_{\mathrm{c}}$ & $w$ & $\chi^2_{\mathrm{norm}}$\\\hdashline
$s$-wave & 1.86(8) &  - & 1  & 1.92  \\ 
$\left(s+s\right)$-wave & 2.3(4)  & 0.3(3) & 0.95(4) & 1.82 \\  
\hline\hline
\end{tabular}

\caption{Superconducting gap parameters for \NRB\ extracted from fits to the temperature dependence of the inverse penetration depth squared $\lambda^{-2}\left(T\right)$ in the clean and the dirty limits.}
\label{MuonFits}
\end{table}

The magnetic penetration depth given by the $s$-wave model is $\lambda(0) = 595(5)$~nm. The coherence length, $\xi$, determined using $\mu_0H_{c2} = \Phi_0/2\pi\xi^2$ can be combined with $\mu_0H_{\mathrm{c1}}$ to provide an independent estimate of the penetration depth $\lambda$ via the relationship $\mu_0H_{\mathrm{c1}} = (\Phi_0/4\pi\lambda^2)\ln\left(\lambda/\xi\right)$.  Using $\mu_0H_{\mathrm{c1}}(0) = 4.6(1)$~mT obtained in Section~\ref{Magnetic} and $\mu_0H_{\mathrm{c2}}(0) \approx 18$~T in Section~\ref{Upper Critical Field} below, we find $\lambda(0) = 403(5)$~nm, which is in reasonable agreement with our estimate of $\lambda(0)$ from the muon spectroscopy data~\footnote{The $H_{\mathrm{c1}}(0)$ determined from the $M$ vs $H$ data may be overestimated, as the values of $H_{\mathrm{c1}}(T)$ used in Figure~\ref{Magnetisation}(c) are recorded when the sample has clearly entered the mixed state. A slightly lower value of $H_{\mathrm{c1}}(0)$ gives a $\lambda(0)$ that matches well with the value determined from the $\mu$SR data}. Both estimates are significantly larger than the $\lambda(0)$ calculated from magnetisation measurements in Ref.~\cite{Carnicom}, where $\lambda(0) = 219(7)$~nm. 

\begin{figure}[tb]
\centering
\includegraphics[width=0.5\columnwidth]{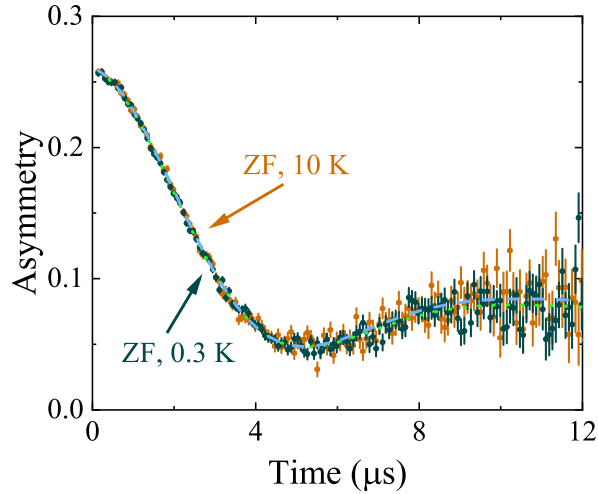}
\caption{(Colour online) Zero-field $\mu$SR spectra of \NRB\ above (10~K) and below (0.3~K) the superconducting transition. No measurable difference in the relaxation of the asymmetry between the two spectra indicates that time-reversal symmetry is preserved in \NRB. The dotted and dashed line shows the fit to the two spectra using a Gaussian Kubo-Toyabe function.}
\label{LF_spectra}
\end{figure}

In order to investigate whether time-reversal symmetry is broken in \NRB, zero-field (ZF) muon spin relaxation spectra were collected in the normal and superconducting state~\cite{LaNiC2, Re6Zra, La7Ir3}. Figure~\ref{LF_spectra} shows the two spectra have a Kubo-Toyabe-like form~\cite{Kubo-Toyabe} with no measurable difference between the two data sets, indicating that time-reversal symmetry is preserved in \NRB. The absence of any oscillatory component or loss of initial asymmetry in the zero-field $\mu$SR spectra shows that no magnetic ordering takes place in \NRB\ down to 300~mK.  

\section{Pressure Dependence of \TC}
\label{Pressure}

\begin{figure}[tb]
\centering
\includegraphics[width=0.5\columnwidth]{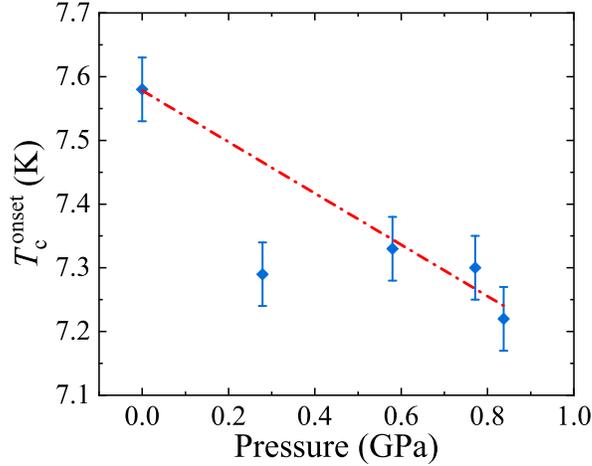}
\caption{(Colour online) Pressure dependence of the superconducting transition temperature, \TC, for \NRB. A small decrease in the transition temperature can be seen for increasing pressure.}
\label{FIG:Pressure}
\end{figure}

The change in \TC\ with hydrostatic pressure is shown in Fig.~\ref{FIG:Pressure}. The shift in the transition temperature was determined from magnetisation versus temperature data collected in an applied field of 1~mT. \TC\ was estimated by extrapolating the superconducting response at the transition slope and taking the intersection with a line drawn from the normal-state susceptibility at higher temperature. The superconducting transition decreases sharply as the initial pressure is applied before decreasing more slowly thereafter. A linear fit to the data gives $\mathrm{d}T_{\mathrm{c}}/\mathrm{d}P = -0.40(4)$~K/GPa. The width of the transition is not seen to broaden noticeably across the available pressure range. A decrease in \TC\ with pressure is typical for simple BCS superconductors, including the elements~\cite{Parks69}. $\mathrm{d}T_{\mathrm{c}}/\mathrm{d}P$ for \NRB\ is the same order of magnitude as that seen in pure Nb [$-0.28\left(2\right)$~K/GPa], Ta [$-0.26\left(1\right)$~K/GPa], and V [$-0.49\left(5\right)$~K/GPa] which are all transition metals with similar isothermal compressibility. $\mathrm{d}T_{\mathrm{c}}/\mathrm{d}P$ normalised to \TC\ is also very close to the value of $\mathrm{d}T_{\mathrm{c}}/\mathrm{d}P=-2.14\left(6\right)$~K/GPa observed for the two-band boride superconductor MgB$_{2}$. Variations in \TC\ with pressure are typically driven by changes in the Debye temperature or the density of states close to the Fermi energy~\cite{Parks69}. Measurements of resistivity under pressure would help to determine the change in $\theta_{\mathrm{D}}$ with pressure, while detailed calculations of the band structure may shed more light on the role played by changes in $N\left(E_{\mathrm{F}}\right)$. Other NCS superconductors also exhibit a decrease in \TC\ with pressure. For example, La$_{7}$Ir$_3$ exhibits a $\mathrm{d}T_{\mathrm{c}}/\mathrm{d}P = -0.15$~K/GPa~\cite{Li18}. As noted above, the \TC\ of Mo$_{3}$Al$_{2}$C increases non-monotonically with $\mathrm{d}T_{\mathrm{c}}/\mathrm{d}P = +0.28(2)$~K/GPa close to \TC~\cite{Bauer10}. The more conventional behaviour seen in \NRB\ suggests that the superconducting gap is dominated by an $s$-wave component.  

\section{Heat Capacity}
\label{Heat Capacity}

\begin{figure}[tb]
\centering
\includegraphics[width=0.5\columnwidth]{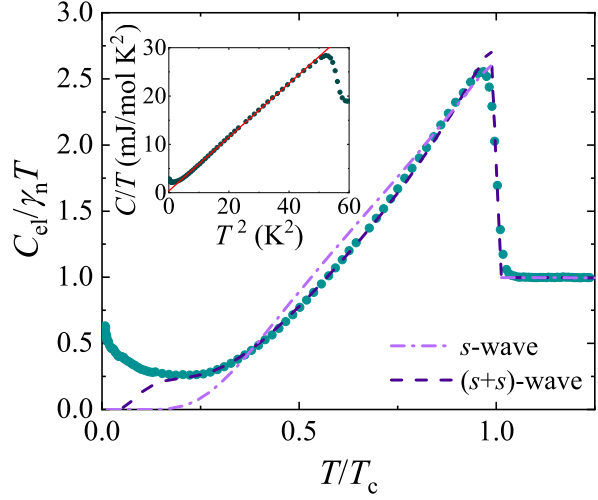}
\caption{(Colour online) Zero-field specific heat of \NRB\ with the phonon contribution subtracted divided by $\gamma_{\mathrm{n}} T$ where $\gamma_{\mathrm{n}}$ is the Sommerfeld coefficient. Fits to the data between \TC\ and 1.5~K are shown using a single-gap isotropic $s$-wave model (light purple) and an isotropic two-gap $\left(s+s\right)$-wave (blue) model. The inset shows that the zero-field specific heat has a $T^{3}$ dependence as demonstrated by the linear fit to specific heat divided by temperature as a function of $T^{2}$.}
\label{SCgap}
\end{figure}

The heat capacity in the normal state contains no anomalies between 300~K and \TC. This confirms that there are no changes in the crystal symmetry on cooling. Immediately above \TC, $C\left(T\right)=\gamma_{\mathrm{n}}T + \beta T^3$  where $\gamma_{\mathrm{n}}$ is the Sommerfeld coefficient. $\gamma_{\mathrm{n}}= 7.6\left(4\right)$~mJ/mol K$^2$ and $\theta_{\mathrm{D}}$, calculated from $\theta_{\mathrm{D}}=\left(\frac{12\pi^4}{5}Nk_{\mathrm{B}}/\beta\right)^{\frac{1}{3}}$, is $388\left(9\right)$~K. In zero-field, the onset of bulk superconductivity is indicated by an anomaly in the specific heat where the midpoint of the jump is defined as the transition temperature. For \NRB, $T_{\mathrm{c}} = 7.46(5)$~K and $\Delta C/ \gamma_{\mathrm{n}} T_{\mathrm{c}} = 1.69$. $\Delta C/ \gamma_{\mathrm{n}} T_{\mathrm{c}}$ is larger than the 1.43 expected for a conventional BCS superconductor, again indicating moderate to strong coupling. The temperature dependence of the electronic specific heat below \TC/10 can be used to establish the nature of the superconducting gap~\cite{Mazidian13}. An exponential behaviour is expected for a conventional nodeless BCS gap, while a power law temperature dependence indicates that there may be nodes in the gap~\cite{Bauer10}. However, as seen in Fig.~\ref{SCgap}, we observe an upturn in $C_{\mathrm{el}}\left(T\right)/\gamma_{\mathrm{n}}T_{\mathrm{n}}$ in \NRB\ at low-temperatures. This upturn is also present in applied magnetic fields up to 1~T but is not observed at 2~T (data not shown). There is no indication of any sizeable magnetic moments present in the sample from the magnetic susceptibility in the normal state or the muon spectroscopy data so the upturn cannot be a precursor due to magnetic ordering. An upturn in the heat capacity at low-temperature may arise from a hyperfine contribution. Niobium and boron have nuclear moments of $\mu/\mu_{\mathrm{N}} = 6.17$ and $\mu/\mu_{\mathrm{N}} = 2.75$ respectively, so a Schottky anomaly shifting with applied field may be expected. A similar feature with the same temperature and field dependence is seen in \TRB~\cite{Mayoh}. In a simple case, any hyperfine contribution to $C\left(T\right)$ is expected to follow a $1/T^{2}$ dependence. The contribution observed at low-temperature in Fig.~\ref{SCgap} does not have this temperature dependence. It is possible to model this upturn using a simple (arbitrary) $1/T^{n}$ temperature dependence, but subtracting such a contribution necessarily leaves the temperature dependence of the electronic component $C_{\mathrm{el}}\left(T\right)$ uncertain. As a result of the upturn in $C\left(T\right)/T$, it is not possible to definitively establish the nature of the superconducting gap for \NRB\ from heat capacity measurements. Nevertheless, the data at higher temperatures allows progress to be made. For the following analysis, the temperature region affected by the anomalous contribution was excluded. The temperature dependence of the zero-field electronic specific heat $C_{\mathrm{el}}$ was fit with a two-gap model~\cite{Bouquet01b}. This phenomenological model postulates the existence of two gaps rather than explaining their origin and assumes a BCS-like temperature dependence for each gap~\cite{Gapfitting}. It has, however, been successfully used to model the heat capacity of a wide range of superconductors. The entropy $S$ was calculated from 

\begin{equation}
\dfrac{S_{i}}{\gamma_{\mathrm{n}} T_{\mathrm{c}}}=-\dfrac{3}{\pi^3}\int^{2\pi}_{0}\int_{0}^{\infty}\left[f\mathrm{ln}f+\left(1-f\right)\mathrm{ln}\left(1-f\right)\right]dyd\phi,
\end{equation}

\noindent where $f$ is the Fermi-Dirac function and $E=\sqrt{y^2+\left[\Delta_{0}\left(T,\phi\right)\right]^{2}}$, where $y$ is the energy of the normal state electrons and $\Delta_{0}\left(T,\phi\right)$ captures the temperature and angular dependence of the superconducting gap function. For $s$-wave superconductivity $\Delta_{0}\left(T,\phi\right)= \Delta_{0}\left(T\right)$. The specific heat of the superconducting state is then calculated by
\begin{equation}
\dfrac{C_{\mathrm{sc}}}{\gamma_{\mathrm{n}} T}=\dfrac{d(S/\gamma_{\mathrm{n}} T_{\mathrm{c}})}{dT}.
\end{equation}

\noindent For the two-gap model the total electronic specific heat can be considered a weighted sum of each gap calculated separately~\cite{Bouquet01, Zehetmayer} given by:
\begin{equation}
C_{\mathrm{el}}\left(T\right) = w C_{\mathrm{sc,}1}\left(T\right) +\left(1-w\right)C_{\mathrm{sc,}2}\left(T\right).
\end{equation}\noindent
The zero-field $C_{\mathrm{el}}\left(T\right)$ data between \TC\ and 1.5~K were fit using a single-gap isotropic $s$-wave model and an isotropic two-gap $\left(s+s\right)$-wave model. The two-gap model produces a better fit to the data (see Fig.~\ref{SCgap}). The gap values for the $\left(s+s\right)$ model are $\Delta_{0,1}/k_{\mathrm{B}}T_{\mathrm{c}} = 2.16(1)$, $\Delta_{0,2}/k_{\mathrm{B}}T_{\mathrm{c}} = 0.39(4)$ with $w = 0.801(3)$. The larger of the two gaps is above the expected BCS value of 1.76 for a weakly-coupled superconductor, indicating a stronger electron-phonon coupling. The small size for the second gap may indicate that the temperature at which this gap closes is raised relative to the normal BCS $T_{\mathrm{c}}$ via coupling to the larger gap. The observation of multigap behaviour in \NRB\ is consistent with previous studies of isostructural \TRB~\cite{Mayoh}.

As seen in the inset of Fig.~\ref{SCgap} below \TC, $C_{\mathrm{el}}\left(T\right)$ deviates from an $s$-wave BCS-like behaviour and instead exhibits a $T^{3}$ dependence, suggesting the presence of a node in the gap or multigap behaviour. This behaviour is also observed in other NCS two-gap superconductors including \TRB~\cite{Mayoh} and Mo$_{3}$Al$_{2}$C~\cite{Mo3Al2C, Bauer10}, as well as several well-studied centrosymmetric superconducting materials such as Lu$_2$Fe$_3$Si$_5$~\cite{Nakajima08, BiswasLu2fe3Si5} and MgB$_2$~\cite{Bouquet01}. This power law dependence and the multigap fit of the heat capacity in the superconducting state are reasonably consistent with the values obtained for the two-gap fit to the muon spectroscopy data, but at odds with the nodeless $s$-wave superconducting ordering parameter discussed in Section~\ref{muSR}. The anomalous low-temperature contribution to the heat capacity, therefore, leaves open the question of the precise nature of the superconducting order parameter in \NRB. 

\section{Upper Critical Field}
\label{Upper Critical Field}

\begin{figure}[tb]
\centering
\includegraphics[width=0.45\columnwidth]{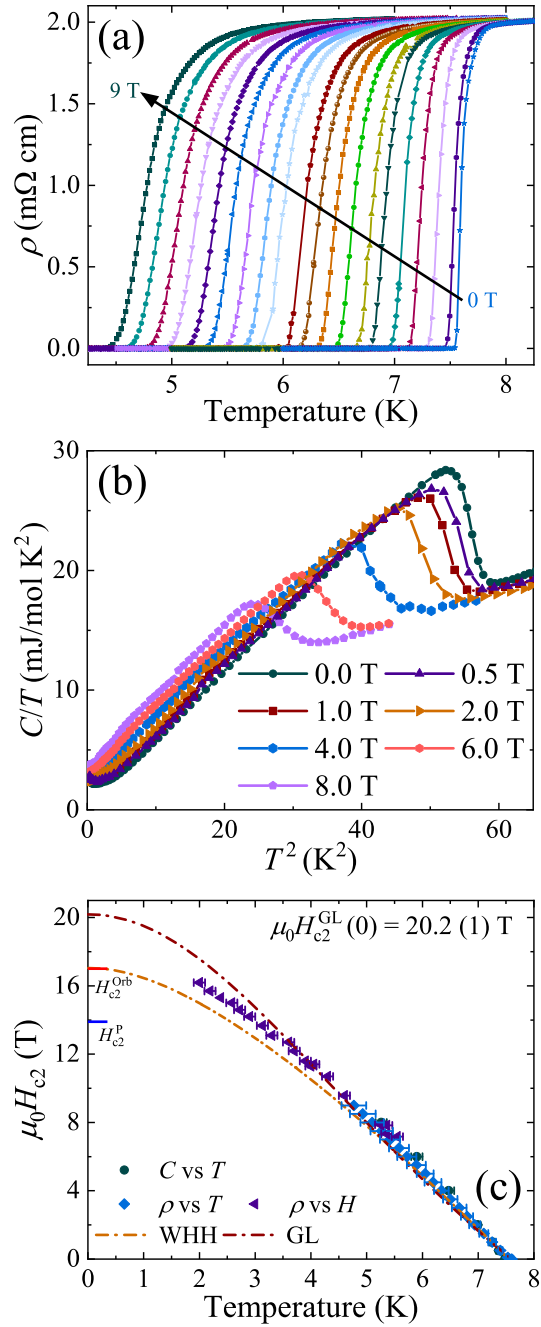}
\caption{(Colour online) (a) Temperature dependence of the electrical resistivity for \NRB\ in applied magnetic fields up to 9~T. (b) Specific heat divided by temperature $C\left(T\right)/T$ as a function of $T^2$ for \NRB\ in various applied magnetic fields. (c) Upper critical field as a function of temperature for \NRB\ where the $H_{\mathrm{c2}}\left(T\right)$ points were extracted from the \TC\ in heat capacity and electrical resistivity as a function temperature and field. Fits using the WHH and GL models are shown by the dashed and the dashed-dotted lines respectively.}
\label{UpperCrit}
\end{figure} 

The upper critical field as function of temperature, $\mu_{0}H_{\mathrm{c2}}\left(T\right)$, in \NRB\ was obtained by measuring \TC\ from heat capacity up to 8~T (see Section~\ref{Heat Capacity}), and resistivity up to 17~T, (see Fig.~\ref{UpperCrit}). \NRB\ is a poor metal. $\rho\left(T\right)$ decreases below room temperature with a minimum in $\rho\left(T\right)$ at around 130~K. An upturn at lower $T$ suggests the carriers then become more localized. $\rho_{300~\mathrm{K}}/\rho_{10~\mathrm{K}}=1.09(1)$ and the resistivity at 10~K, just above \TC, is $\sim 2~\mathrm{m}\Omega$~cm. This rather high value may be due to a combination of strong electron-phonon scattering, disorder, and poor connectivity between the grains of the polycrystalline samples. Spin-orbit scattering may also play a role. In Ta$_{2}$Pd$_{x}$S$_{5}$, it is suggested that strong spin-orbit scattering due to Pd deficiencies gives rise to a substantial increase in $\mu_{0}H_{\mathrm{c2}}\left(0\right)$~\cite{Lu14}. 

The temperature dependence of $\mu_{0}H_{\mathrm{c2}}$ determined from the resistivity and heat capacity data is almost linear up to 9~T. $\mu_{0}H_{\mathrm{c2}}\left(T\right)$ with a linear temperature dependence is often fit using the Werthamer-Helfand-Hohenberg model, which allows $\mu_{0}H_{\mathrm{c2}}\left(T\right)$ to be calculated taking into account both spin-orbit scattering and Pauli spin paramagnetism~\cite{WHH66}. Niobium and rhodium are both heavier elements where spin-orbit coupling may be expected to be significant and therefore to have an influence on the value of $\mu_{0}H_{\mathrm{c2}}$. The WHH theory gives

\begin{eqnarray}
\ln\left(\dfrac{1}{t}\right) &= \left(\dfrac{1}{2} + \dfrac{i\lambda_{\mathrm{so}}}{4\gamma}\right)\psi\left(\dfrac{1}{2}+\dfrac{h+\frac{1}{2}\lambda_{\mathrm{so}}+i\gamma}{2t}\right) \nonumber\\
&+\left(\dfrac{1}{2}-\dfrac{i\lambda_{\mathrm{so}}}{4\gamma}\right)\psi\left(\dfrac{1}{2}+\dfrac{h+\frac{1}{2}\lambda_{\mathrm{so}}+i\gamma}{2t}\right)-\psi\left(\dfrac{1}{2}\right),
\end{eqnarray}

\noindent where $\psi$ is the digamma function, $t = T/T_{\mathrm{c}}$, $\lambda_{\mathrm{so}}$ is the spin-orbit scattering parameter, $h$ is the dimensionless form of the upper critical field given by
\begin{equation}
h = \dfrac{4 H_{\mathrm{c}2}}{\pi^2}\left.\left(\dfrac{\mathrm{d}H_{\mathrm{c}2}}{\mathrm{d}t}\right)^{-1}\right|_{t=1},
\end{equation}

\noindent $\gamma=\sqrt{(\alpha_{\mathrm{M}} h)^2-(\frac{1}{2} \lambda_{\mathrm{so}})^2}$, and $\alpha_{\mathrm{M}} \left[= \sqrt{2}H_{\mathrm{c2}}^{\mathrm{orb}}\left(0\right)/H_{\mathrm{P}}\right]$ is the Maki parameter. Initially $\alpha_{\mathrm{M}}$ was estimated to be $\sim 1.8$, but in the subsequent fits both $\alpha_{\mathrm{M}}$ and $\lambda_{\mathrm{so}}$ were allowed to vary and this resulted in $\alpha_{\mathrm{M}}\rightarrow 0$ and values for $\lambda_{\mathrm{so}}>>1$, well above those expected for the WHH formalism~\cite{WHH66, Clogston62}.  The fit, shown with the dot-dash line, produces the maximum value for the upper critical field, $\mu_0H_{\mathrm{c2}}^{\mathrm{orb}}\left(0\right)\approx 17$~T that is consistent with the gradient of the data, $dH_{\mathrm{c2}}\left(T\right)/dT$, close to \TC. This value for $H_{\mathrm{c2}}\left(T\right)$ is comparable to that reported in Ref.~\cite{Carnicom} and well above the Pauli paramagnetic limiting field $\left(\mu_{0}H_{\mathrm{c2}}^{\mathrm{P}}\left[\mathrm{T}\right] = 1.85T_{\mathrm{c}}\left[\mathrm{K}\right]\right)$ of $13.9\left(2\right)$~T. However, the data clearly deviate from the WHH model at lower $T$ and the WHH model does not provide a satisfactory description of the data. We have also used the phenomenological Ginzburg-Landau (GL) expression

\begin{equation}
H_{\mathrm{c2}}\left(T\right)=H_{\mathrm{c2}}\left(0\right)\left[1-\left(T/T_{\mathrm{c}}\right)^2\right]/\left[1+\left(T/T_{\mathrm{c}}\right)^2\right],
\end{equation}

\noindent to estimate $\mu_0H_{\mathrm{c2}}\left(0\right)=20.2(2)$~T, but this is clearly higher than the value that would be estimated from a visual inspection of the $H_{\mathrm{c2}}\left(T\right)$ data. Neither model can fully explain the temperature dependence of the upper critical field of \NRB\ suggesting that it is necessary to consider two-band effects to describe the behaviour. A model proposed by Gurevich that considers intra- and inter-band scattering has been used to successfully describe the $T$ dependence of $H_{\mathrm{c2}}$ of several two-band superconductors~\cite{Gurevich}. The model includes several free parameters, so any meaningful consideration of $H_{\mathrm{c2}}\left(T\right)$ for \NRB\ within this framework requires a more complete knowledge of the intra- and inter-band scattering. Such an analysis needs single crystals. While multigap behaviour may explain why the upper critical field of \NRB\ exceeds the Pauli limit, it is known that strong pinning and energy gaps larger than the BCS value can also be responsible for increasing the Pauli limiting field~\cite{Clogston62, Orlando79}. The magnetisation and susceptibility data, however, indicate the pinning is rather weak in \NRB. The magnitude of the superconducting energy gap is larger than the BCS value, but an enhancement of this kind can only increase the Pauli limit by around 30\%~\cite{Clogston62}.

\section{Summary}

In summary, we have investigated the superconducting properties of the chiral noncentrosymmetric superconductor \NRB\ using a combination of $\mu$SR, magnetisation, heat capacity, and resistivity measurements. We have shown that \NRB\ is a bulk type-II superconductor with a superconducting transition at $7.46(5)$~K that is mediated by moderate electron-phonon coupling. The temperature dependence of the penetration depth measured by TF $\mu$SR indicates that \NRB\ has a superconducting gap that can be fit by both an $s$-wave or an $\left(s + s\right)$-wave model.  The magnetic penetration depth is $\lambda(0) = 595(5)$~nm. The superconducting gap measured by heat capacity data, although convoluted with an anomalous low-temperature contribution, strongly indicates a multigap $\left(s + s\right)$-wave dependency with a conventional $s$-wave model providing a poor fit to the data. The decrease in \TC\ with pressure suggests that the superconducting gap is dominated by an $s$-wave component. We have also shown that the upper critical field of \NRB\ at low temperatures is not adequately described by either the WHH or GL models and that a two-band formalism may be required to more fully account for the $\mu_{0}H_{\mathrm{c2}}\left(0\right)$ behaviour. Strong spin-orbit scattering may also play an important role in the physics of \NRB\ and more complete knowledge of the intra- and inter-band scattering, including scattering due to non-magnetic defects, is required to complete such an analysis. To gain a deeper understanding of the enhancement in $\mu_{0}H_{\mathrm{c2}}\left(0\right)$ and the nature of superconducting gap in \NRB\ (and \TRB), high-quality single crystals are urgently required. Tunnel-diode oscillator measurements of the penetration depth may also help shed light on the form of the superconducting gap in this interesting new family of superconductors.

\pagebreak

\subsection{Acknowledgements}
This work is funded by the EPSRC, United Kingdom, through grant EP/M028771/1. This project has received funding from the European Research Council (ERC) under the European Union’s Horizon 2020 research and innovation programme (Grant agreement No. 681260).

\section*{References}
\bibliography{NbRh2B2_DM_References}

\end{document}